\documentclass[twocolumn,aps,showpacs]{revtex4}

\def\xsxnxnpp{$\sigma(AuAu \!\rightarrow\! Au^*_{xn} Au^*_{xn} \rho^0)\!= 28.3 \!\pm\! 2.0 \pm 6.3$~mb}
\def\xssnsn{$\sigma(AuAu \!\rightarrow\! Au^*_{1n} Au^*_{1n} \rho^0) \!=\! 2.8 \!\pm\! 0.5\!\pm\! 0.7$~mb}
\def\xsxnxn{$\sigma^{{\rm (inc. overlap)}}(AuAu \!\rightarrow\! Au^*_{xn} Au^*_{xn} \rho^0) \!=\!  39.7  \!\pm\! 2.8  \!\pm\! 9.7$~mb}
\def\xsnobrk{$\sigma(AuAu \!\rightarrow\! Au Au \rho^0) \!=\!  370  \!\pm\! 170 \pm 80$~mb}
\def\xsxnzn{$\sigma(AuAu \!\rightarrow\! Au^\star_{xn} Au \rho^0) \!=\!  95 \!\pm\! 60 \pm 25$~mb}
\def\xstot{$\sigma(AuAu \!\rightarrow\! Au^{(*)} Au^{(*)} \rho^0) \!=\!  460  \!\pm\! 220 \pm 110$~mb}
\def\Lumi{$L\!=\!59$~mb$^{-1}$}

\usepackage{graphicx}% Include figure files
\usepackage{dcolumn} % Align table columns on decimal point
\usepackage{bm}      % bold math
\begin{document}

\title{Coherent $\mathbf{ \rho^0}$ Production in Ultra-Peripheral Heavy Ion Collisions }

\author{
C.~Adler$^{11}$, Z.~Ahammed$^{23}$, C.~Allgower$^{12}$, J.~Amonett$^{14}$,
B.D.~Anderson$^{14}$, M.~Anderson$^5$, G.S.~Averichev$^{9}$, 
J.~Balewski$^{12}$, O.~Barannikova$^{9,23}$, L.S.~Barnby$^{14}$
% ...
\\(STAR Collaboration)
\\
% ...
$^5$University of California, Davis, California 95616
\\
% ...
$^{9}$Laboratory for High Energy (JINR), Dubna, Russia
\\
% ...
$^{11}$University of Frankfurt, Frankfurt, Germany
\\$^{12}$Indiana University, Bloomington, Indiana 47408
\\
% ...
$^{14}$Kent State University, Kent, Ohio 44242
\\
% ...
$^{23}$Purdue University, West Lafayette, Indiana 47907
\\
% ...
}

\break
\begin{abstract}
\vskip -.2 in
The STAR collaboration reports the first observation of exclusive
$\rho^0$ photo-production, $AuAu \!\rightarrow \!AuAu \rho^0$, and $\rho^0$
production accompanied by mutual nuclear Coulomb excitation, $AuAu
\!\rightarrow \!Au^\star Au^\star \rho^0$, in ultra-peripheral heavy-ion
collisions.
The $\rho^0$ have low transverse momenta, consistent with coherent coupling to both nuclei. 
The  cross sections  at $\sqrt{s_{NN}}\!=\!130$~GeV agree with theoretical predictions
treating $\rho^0$ production and Coulomb excitation as independent processes. 
\end{abstract}
\maketitle

In ultra-peripheral heavy-ion collisions the two nuclei geometrically
`miss' each other and no hadronic nucleon-nucleon collisions occur. At
impact parameters $b$ significantly larger than twice the nuclear
radius $R_A$, the nuclei interact by photon exchange and photon-photon
or photon-Pomeron collisions~\cite{baurrev}.  Examples are nuclear
Coulomb excitation, electron-positron pair and meson production, and
vector meson production.  The exchange bosons can couple coherently to
the nuclei, yielding large cross sections. Coherence restricts the
final states to low transverse momenta, a distinctive experimental
signature.  The STAR collaboration reports the first observation of
coherent exclusive $\rho^0$ photo-production, $AuAu \!\rightarrow
\!AuAu \rho^0$, and coherent $\rho^0$ production accompanied by mutual
nuclear excitation, $AuAu \!\rightarrow \!Au^\star Au^\star
\rho^0$. Ultra-peripheral heavy-ion collisions are a new
laboratory for diffractive interactions, complementary to fixed-target
$\rho^0$ photo-production on complex nuclei~\cite{alvensleben}.
\begin{figure}[!b]
\includegraphics[width=7.5cm,height=1.8cm,bbllx=0pt,bblly=0pt,bburx=650pt,bbury=180pt]{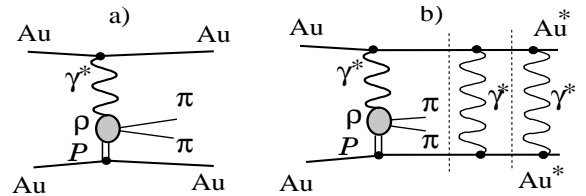}
\caption[]{ Diagram for (a) exclusive $\rho^0$ production in
ultra-peripheral heavy ion collisions, and (b) $\rho^0$ production
with  nuclear excitation.
The dashed lines indicate factorization. 
\label{fig:feynman}}
\end{figure}

Exclusive $\rho^0$ meson production, $AuAu\!  \rightarrow\! Au
Au \rho^0$ (c.f. Fig.~\ref{fig:feynman}a), can be described by the
Weizs\"acker-Williams approach~\cite{weizsaecker} to the photon flux
and the vector meson dominance model~\cite{sakurai}.
A photon emitted by one nucleus fluctuates to a virtual $\rho^0$ meson, 
which scatters elastically from the other
nucleus. The gold nuclei are not
disrupted, and the final state consists solely of the two nuclei and
the vector meson decay products~\cite{BKN}.
In the rest frame of the target nucleus, 
mid-rapidity $\rho^0$ production at RHIC  corresponds to  a  photon energy of
$50$~GeV and a photon-nucleon center-of-mass energy of
$10$~GeV. At this energy, Pomeron $(\cal{P})$ exchange dominates over
meson exchange, as indicated by the rise of the $\rho^0$ production cross section with increasing 
energy in lepton-nucleon scattering~\cite{Crittenden}.
In addition to coherent $\rho^0$ production, the exchange of virtual
photons may excite the nuclei. These  processes are assumed 
to factorize for heavy-ion collisions, which is  justified by the similar case of 
two-photon interactions in  relativistic ion collisions  accompanied by
nuclear breakup, where it was shown that  the non-factorisable diagrams are
small~\cite{hencken}. The process $AuAu
\!\rightarrow\! Au^\star Au^\star \rho^0$ is shown in
Fig.~\ref{fig:feynman}b.  In lowest order, mutual nuclear excitation
of heavy ions occurs by the exchange of two photons~\cite{xsectAuAu,mutualbreakup}.
Because of the Coulomb barrier for the emission of charged particles,
nearly all nuclear decays following photon absorption include neutron
emission~\cite{GDR}.

The photon and Pomeron can couple coherently to the gold nuclei.
The wavelength $\lambda_{\gamma,\cal{P}}\!>\!2R_A$ leads
to coherence conditions: a low transverse momentum of $p_T \!<\! \pi
\hbar/ R_A$ ($\!\sim\!90$~MeV/c for gold with $R_A \!\sim\! 7$~fm),
and a maximum longitudinal momentum of $p_\|\!< \! \pi \hbar \gamma /
R_A$ ($\!\sim\!6$~GeV/c at $\gamma\!=\!70$), where $\gamma$ is the
Lorentz boost of the nucleus.
The $\rho^0$ production cross sections are large. The photon flux is
proportional to the square of the nuclear charge
$Z^2$~\cite{weizsaecker}, and the forward cross section for
elastic $\rho^0 A$ scattering $d\sigma^{\rho A}/dt|_{t=0}$ scales as $A^{4/3}$ for surface
coupling and $A^2$ in the bulk limit.  At a center-of-mass energy of
$\sqrt{s_{NN}}\!=\!130$~GeV per nucleon-nucleon pair, a
total $\rho^0$ cross section, regardless of nuclear excitation,
$\sigma(AuAu
\!\rightarrow\!Au^{(\star)}Au^{(\star)}\rho^0)\!=\!350$~mb is
predicted from a Glauber extrapolation of $\gamma p \!\rightarrow \! 
\rho^0 p$ data~\cite{BKN}.  Calculations for coherent $\rho^0$
production with nuclear excitation assume that both
processes are independent, sharing only a common impact
parameter~\cite{xsectAuAu,BKN}.

In the year 2000, the Relativistic Heavy Ion Collider (RHIC) at
Brookhaven National Laboratory collided gold nuclei at
$\sqrt{s_{NN}}\!=\! 130$~GeV.  In the Solenoidal Tracker at RHIC
(STAR)~\cite{ackermann}, charged particles are reconstructed with a
cylindrical time projection chamber (TPC)~\cite{TPC} operated in a 0.25~T solenoidal magnetic
field. A central trigger barrel (CTB) of 240 scintillator slats
surrounds the TPC. Two zero degree hadron calorimeters (ZDCs) at
$\pm$\,18~m from the interaction point are sensitive to the neutral
remnants of nuclear break-up, with $98\!\pm\!2\%$ acceptance for
neutrons from nuclear break-up through Coulomb
excitation~\cite{ZDC,mutualbreakup}.

Exclusive $\rho^0$ production has a distinctive signature: the
$\pi^+\pi^-$ from the $\rho^0$ decay in an otherwise `empty'
detector. The tracks are approximately back-to-back in the transverse
plane due to the small $p_T$ of the pair. The gold nuclei remain
undetected within the beam.

Two data sets are used in this analysis.  For $AuAu\! \rightarrow\! 
AuAu \rho^0$, about 30,000 events were collected using a
low-multiplicity `topology' trigger.  The CTB was divided in four
azimuthal quadrants. Single hits were required in the opposite side
quadrants; the top and bottom quadrants acted as vetoes to suppress
cosmic rays.  A fast on-line reconstruction~\cite{L3} removed events
without reconstructible tracks from the data stream.  To study $Au Au
\!\rightarrow\! Au^\star Au^\star \rho^0$, a data set of about 800,000
`minimum bias' events, which required coincident detection of neutrons
in both ZDCs as a trigger, is used.

Events are selected with exactly two oppositely charged tracks forming
a common vertex within the interaction region. The $\rho^0$ candidates
are accepted within a rapidity range $|y_\rho|\!<\!1$.  A systematic
uncertainty of $5\%$ is assigned to the number of $\rho^0$ candidates
by varying the event selection criteria. The specific energy loss
$dE/dx$ in the TPC shows that the event sample is dominated by pion
pairs. Without the ZDC requirement in the topology trigger, cosmic rays are a major
background.  They are removed by requiring that the two pion tracks
have an opening angle of less than 3 radians. Using the energy
deposits in the ZDCs, we select events with at least one neutron
(xn,xn), exactly one neutron (1n,1n), or no neutrons (0n,0n) in each
ZDC, and events with at least one neutron in exactly one ZDC (xn,0n);
the latter two occur only in the topology trigger. A $10\%$
uncertainty arises from the selection of single neutron signals.
\begin{figure}[!t]
 \includegraphics[width=4.2cm,height=3.2cm,bbllx=20pt,bblly=20pt,bburx=570pt,bbury=500pt]{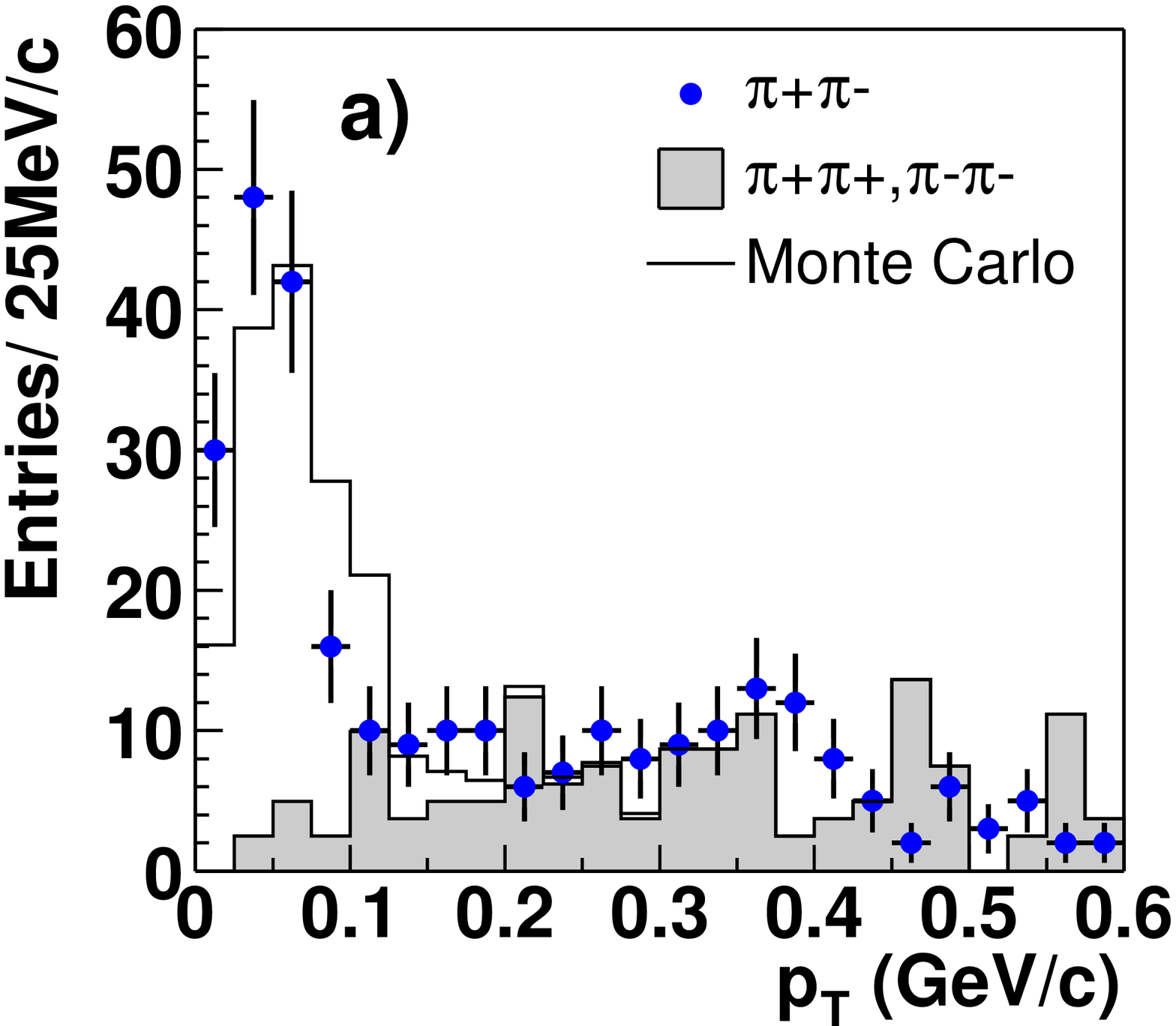}
 \includegraphics[width=4.2cm,height=3.2cm,bbllx=20pt,bblly=20pt,bburx=570pt,bbury=500pt]{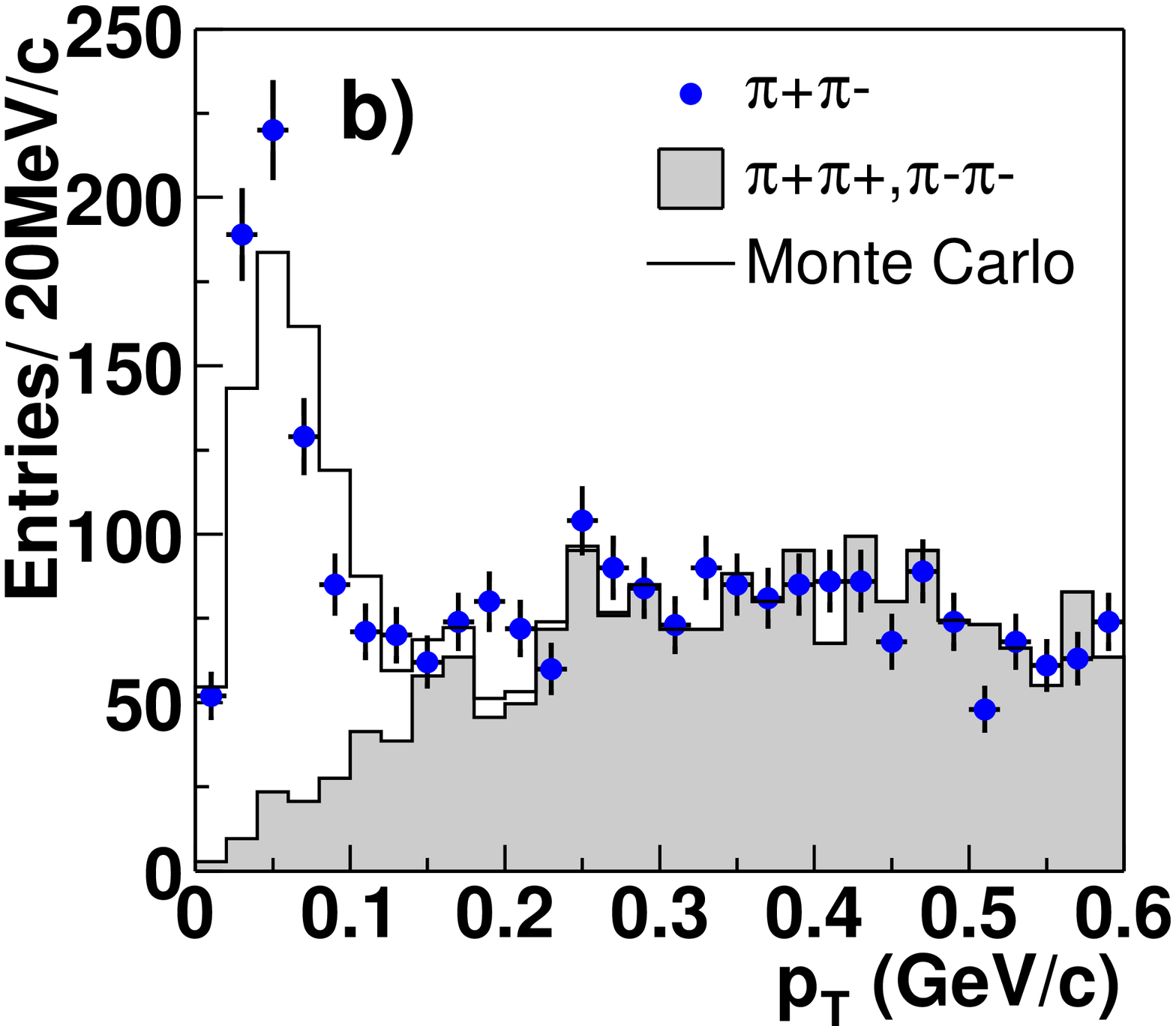}
\caption[]{
The $p_T$ spectra of pion pairs for the 2-track events selected by (a)
the topology trigger (0n,0n) and (b) the minimum bias trigger (xn,xn).
Points are oppositely charged pairs, and the shaded histograms are the
normalized like-sign combinatorial background.  The open histograms
are simulated $\rho^0$ superimposed onto the background. \label{fig:pt}}
\end{figure}

\begin{figure}[!t] 
 \includegraphics[width=8.cm,height=3.2cm,bbllx=20pt,bblly=30pt,bburx=570pt,bbury=280pt]{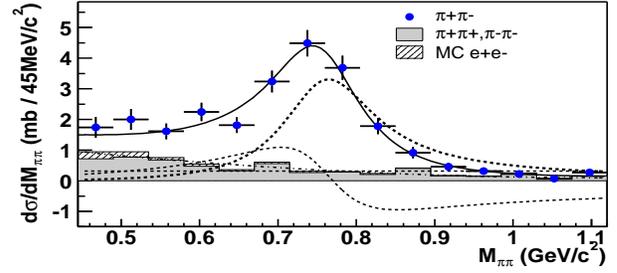}
\caption[]{
The $d\sigma(AuAu \!\rightarrow\! Au^*Au^*\rho)/dM_{\pi\pi}$ spectrum for 2-track (xn,xn) events with
pair-$p_T\!< \!150$~MeV/c in the minimum bias data. The shaded
histogram is the combinatorial background, and the hatched histogram
contains an additional contribution from coherent $e^+e^-$ pairs.  The
fits correspond to Eq.~\ref{eq:bwsoedmod}:
the sum (solid) of a Breit-Wigner, a mass--independent
contribution from direct $\pi^+\pi^-$ production and their
interference (all dashed), and a second order polynomial for  the 
residual background (dash-dotted).
\label{fig:minv}}
\end{figure}

The uncorrected transverse momentum spectra of pion pairs for the two-track event
samples of the topology trigger (0n,0n) and the minimum bias trigger
(xn,xn) are shown in Fig.~\ref{fig:pt}.  Both spectra are peaked at
$p_T\!\sim\!50$~MeV/c, as expected for coherent coupling. A background
model from like-sign combination pairs,  normalized to the
signal at $ p_T \!>\!$~200 MeV/c, is not peaked. For comparison, the
$p_T$ spectra from Monte Carlo simulations~\cite{BKN} discussed below are shown.  They
are normalized to the $\rho^0$ signal at $p_T\!<\!150$~MeV/c and added
to the background.  The $M_{\pi\pi}$ invariant mass spectra
(c.f. Fig.~\ref{fig:minv}) for both event samples are peaked around
the $\rho^0$ mass.  We find $131\pm14$ (0n,0n) and $656\pm36$ (xn,xn)
events at $p_T\!<\!150$~MeV/c, which we define as coherent $\rho^0$
candidates. 

The data contain combinatorial background contributions from grazing
nuclear collisions and incoherent photon-nucleon interactions, which
are statistically subtracted. 
Incoherent $\rho^0$
production, where a photon interacts with a single nucleon, yields
high $p_T$ $\rho^0$s, which are suppressed by the low pair $p_T$
requirement; the remaining small contribution is indistinguishable
from the coherent process.  
A coherently produced background arises
from the mis-identified two-photon process $AuAu\! \rightarrow\! 
Au^{(\star)} Au^{(\star)} l^+l^-$. It contributes mainly at low
invariant mass $M_{\pi\pi} \!<\! 0.5$~GeV/c$^2$.  Electrons with
momenta $p\!<\!140$~MeV/c can be identified by their energy loss
$dE/dx$.  About 30 $e^+e^-$ pairs, peaked at low pair $p_T\! 
\sim\!20$~MeV/c, were detected in the minimum bias data
sample~\cite{Photon2001}. They are extrapolated to the full phase
space using a Monte-Carlo simulation that describes $e^+e^-$ pair
production by lowest order perturbation
theory~\cite{eeMC}. Electron-positron pairs contribute
$4\pm1\%$ to the signal at $p_T\!<\! 150$~MeV/c and $M_\rho\pm
0.3$~GeV/c.  For a given $M_{ll}$, muons have lower momenta than
the corresponding electrons and are less likely to be detected. Their
$<2\%$ contribution to the coherent signal, as well as the
contribution from $\omega$ decays are neglected.

The acceptance and reconstruction efficiency were studied using a
Monte Carlo event generator that reproduces the expected kinematic and
angular distributions for $\rho^0$ production with and without nuclear
excitation~\cite{BKN,lund}, coupled with a full detector simulation.
The $\rho^0$ decay angle distribution is consistent with s-channel
helicity conservation. The $\rho^0$ production angles are not
reconstructed since the $AuAu$ scattering plane can not be determined.
The efficiencies are almost independent of $p_T$ and the reconstructed
invariant mass $M_{\pi\pi}$.  For the minimum bias trigger,
$42\!\pm\!5\%$ of all $\rho^0$ within $|y_\rho|\!<\!1$ are
reconstructed. The topology trigger vetoes the top and bottom of the
TPC, reducing the geometrical acceptance. Pions with
$p_T\!<\!100$~MeV/c do not reach the CTB, effectively excluding pairs
with $M_{\pi\pi}\!<\!500$~MeV/c$^2$. Only $7\!\pm\!1\%$ of all
$\rho^0$ with $|y_\rho|\!<\!1$ are reconstructed in the topology
trigger.  The $p_T$ resolution is 9~MeV/c.  The $M_{\pi\pi}$ and
rapidity resolutions are $11$~MeV/c$^2$ and $0.01$.

\begin{figure}[!t]
\includegraphics[width=4.2cm,height=3.2cm,bbllx=20pt,bblly=40pt,bburx=570pt,bbury=480pt]{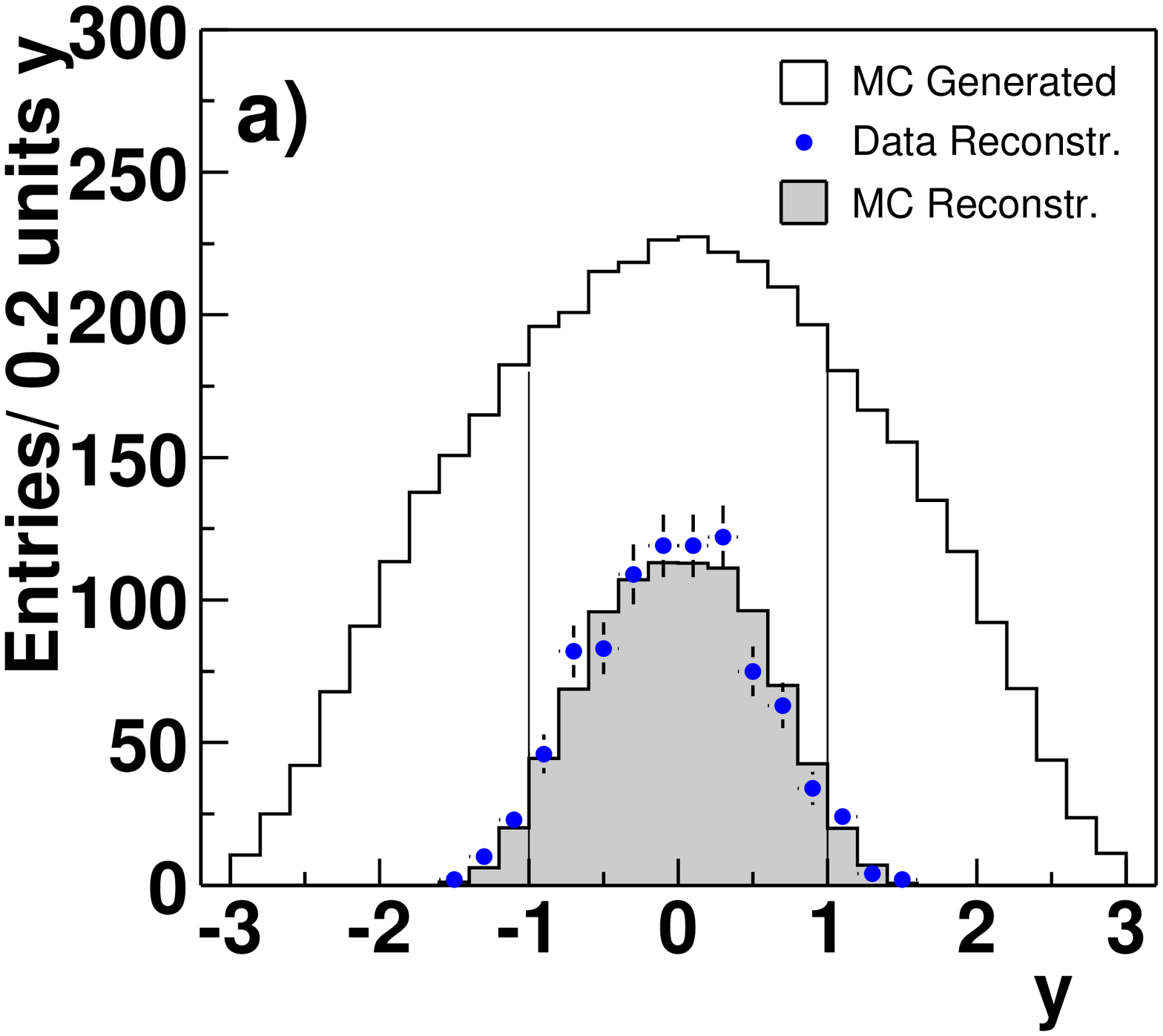}
\includegraphics[width=4.2cm,height=3.2cm,bbllx=20pt,bblly=40pt,bburx=570pt,bbury=480pt]{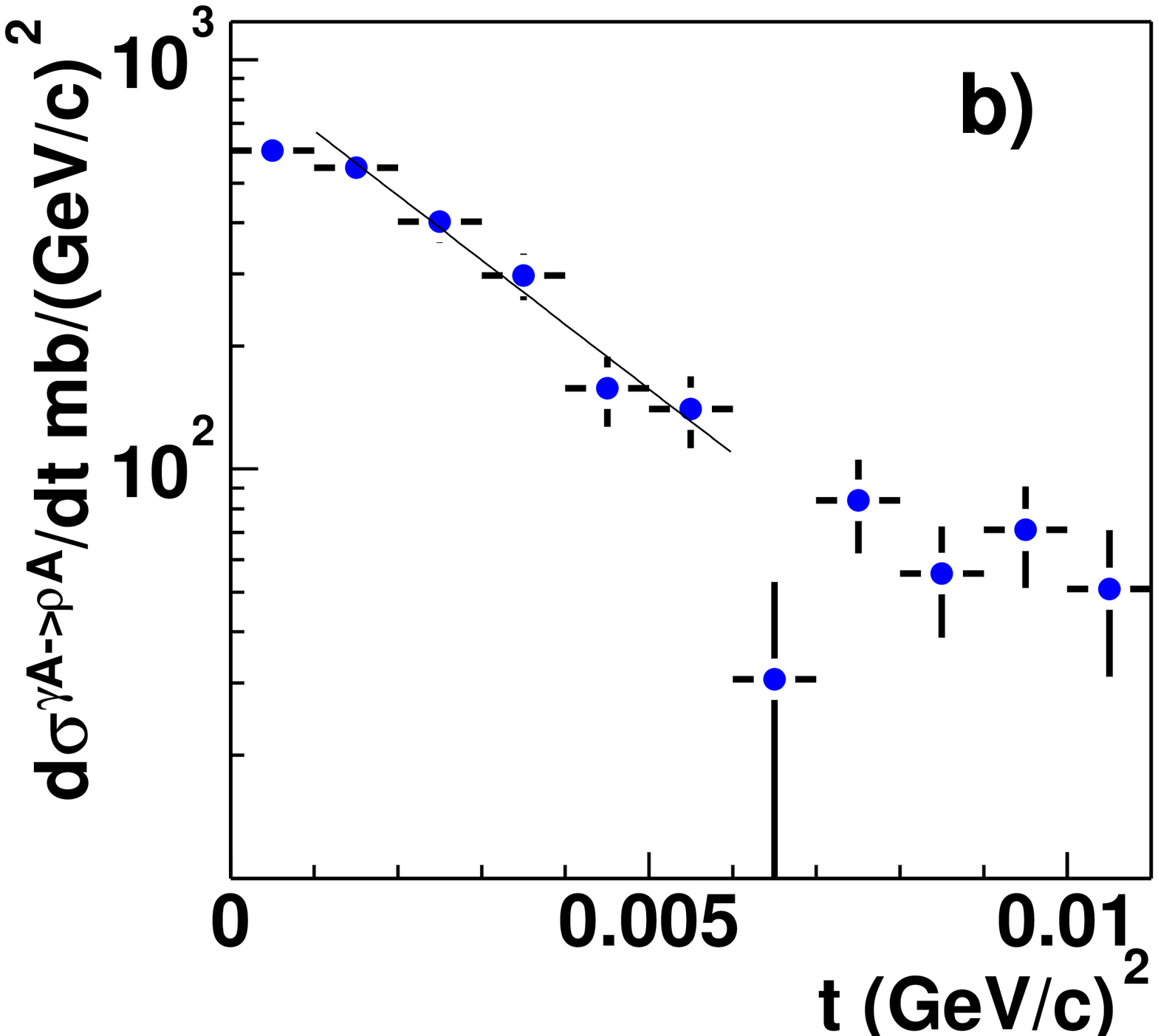}
\caption[]{   
Rapidity distribution (a) of $\rho^0$ candidates (xn,xn) for the minimum
bias data (points) compared to the normalized reconstructed (shaded
histogram) and generated (open histogram) events from the Monte Carlo
simulation. The differential cross section (b)
$d\sigma(\gamma\!Au\!\rightarrow\!\rho\!Au)/dt$ for the same data set; the line
indicates the exponential fit.
\label{fig:rapidity}}
\end{figure}

The rapidity distribution for $\rho^0$ candidates (xn,xn) from the
minimum bias data is shown in Fig.~\ref{fig:rapidity}a).  It is well
described by the reconstructed events from a simulation, which
includes nuclear excitation~\cite{BKN}. The generated rapidity
distribution is also shown.  The acceptance is small for
$|y_\rho|\!>\!1$, so this region is excluded from the analysis.  Cross
sections are extrapolated from $|y_\rho|\!<\!1$ to the full $4\pi$
acceptance by
$\sigma^\rho_{4\pi}/\sigma^\rho_{|y_\rho|\!<\!1}\!=\!1.9$ for $\rho^0$
production with nuclear break-up, and
$\sigma^\rho_{4\pi}/\sigma^\rho_{|y_\rho|\!<\!1}\!=\!2.7$ for
$\rho^0$ production without nuclear break-up. A $15\%$ uncertainty in
the extrapolations is estimated by varying the Monte Carlo parameters.
Event rapidity and photon energy are related by $y=(1/2)
\ln{(2E_\gamma/M_\rho)}$.  But, the average photon
energy  per rapidity bin $\langle E_\gamma\rangle\!\sim \!50$~GeV  is constant, 
when taking the ambiguity of
photon emitter and scattering target into account.

The minimum bias data sample has an integrated luminosity of \Lumi.
The luminosity was measured by counting events containing more than 5
negatively charged hadrons with $p_T\!>\!100$~MeV/c and
pseudo-rapidity $|\eta|\!<\!0.5$. These events represent $79\%$ of the
hadronic cross section~\cite{hadrons}. We assume a total gold-gold
hadronic cross section of $7.2$~b~\cite{xsectAuAu}; its uncertainty
dominates the $10\%$ systematic uncertainty of $L$.

The differential cross section $d\sigma(\gamma Au\!\rightarrow\!\rho Au)/dt \!\sim
\!d\sigma(\gamma Au\!\rightarrow\!\rho Au)/dp_T^2$ for the (xn,xn) events 
is shown in Fig.~\ref{fig:rapidity}b). Here, the combinatorial
background is subtracted. The photon flux is determined by
integration of the photon-spectrum of a relativistic nucleus over the
impact parameter space~\cite{BKN}. For $\rho^0$ production in
ultra-peripheral collisions, $d\sigma/dt$ reflects not
only the nuclear form factor, but also the photon $p_T$ distribution
and the interference of production amplitudes from both gold
nuclei. The interference arises since both nuclei can be either the
photon source or the scattering target~\cite{vminterf}. A detailed
study of this effect is beyond the scope of this paper and the available
statistics. From a fit to $d\sigma^{\rho Au}/dt \propto e^{-bt}$ we
obtain a forward cross section $d\sigma^{\rho
A}/dt|_{t=0}\!=\!965\pm140\!\pm\!230$ mb/GeV$^2$ and an approximate gold radius
of $R_{Au}\!=\!\sqrt{4b}\!=\!7.5\!\pm\!2$~fm, comparable to previous
results~\cite{alvensleben}.

The $d\sigma(AuAu \!\rightarrow\! Au^*Au^*\rho)/dM_{\pi\pi}$ invariant
mass spectrum for the (xn,xn) events with a pair $p_T \!<\!150$~MeV/c
is shown in Fig.~\ref{fig:minv}; the (0n,0n) events have a similar
$d\sigma/dM_{\pi\pi}$ spectrum. Three different parameterizations are
applied:
\begin{eqnarray}
{d\sigma}/{dM_{\pi\pi}} &=& f_\rho \cdot BW(M_{\pi\pi}) + f_{I} \cdot I(M_{\pi\pi}) + f_p,
\label{eq:bwsoed} \\
{d\sigma}/{dM_{\pi\pi}}& =& \bigg| A\frac{ \sqrt{M_{\pi\pi}M_\rho
\Gamma_\rho}}{ M_{\pi\pi}^2 - M_\rho^2 +iM_\rho\Gamma_\rho} + B
\bigg|^2 + f_p,
\label{eq:bwsoedmod} \\
{d\sigma}/{dM_{\pi\pi}} &=& f_\rho \cdot BW(M_{\pi\pi}) \cdot \left({m_\rho}/{M_{\pi\pi}}\right)^n  + f_{p}.
\label{eq:bwrs}
\end{eqnarray}
Eq.~\ref{eq:bwsoed} is a relativistic Breit-Wigner, $BW\! =\! {M_{\pi\pi} M_\rho
\Gamma_\rho}/$ ${ [(M^2_\rho\!-\!M^2_{\pi\pi})^2 \!+\! 
M^2_\rho\Gamma^2_\rho]}$, for $\rho^0$ production plus a S\"oding
interference term~\cite{soding}, $I(M_{\pi\pi}) \!= \!(M^2_\rho\! -\! 
M^2_{\pi\pi})/[(M^2_\rho\!-\!M^2_{\pi\pi})^2 \!+\!
M^2_\rho\Gamma^2_\rho]$, Eq.~\ref{eq:bwsoedmod} is a modified
S\"oding parametrization~\cite{ZEUS}, and Eq.~\ref{eq:bwrs} is a
phenomenological Ross-Stodolsky parametrization ~\cite{rosssto}. Here,
$\Gamma_\rho\! =\! \Gamma_0 \cdot ({M_{\rho}}/{M_{\pi\pi}})\cdot
[(M^2_{\pi\pi} \!-\! 4m^2_\pi)/(M^2_{\rho}\! -\! 4m^2_\pi)]^{{3}/{2}}$
is the momentum-dependent width, and $f_p$ is a fixed second order
polynomial describing the residual background.  The fit parameters are
given in Table~\ref{table:fits}.  The $\rho^0$ mass and width are
consistent with accepted values~\cite{PDG}; they were fixed to reduce
the number of degrees of freedom to obtain $|B/A|, f_I/f_\rho$, and
$n$.  Our results are consistent with values found for the same
parameterizations in $\gamma p\!\rightarrow\! \rho^0 p$
photo-production data~\cite{ZEUS,H1}.
\begin{table}[!t]
\begin{tabular}{|l|c|c|c|}
\hline 
Eq. & $\!M_\rho${\tiny(MeV/c$^2$)}$\!$&$\!\Gamma_\rho^0${\tiny(MeV/c$^2$)}$\!$& \\  
\hline
1 & $778\pm7$ & $148\pm14$  & $f_I/f_\rho \!= \!0.47\!\pm\!0.07\!\pm\!0.12$GeV \\
2 & $777\pm7$ & $139\pm13$  & $|B\!/\!A|\!=\!0.81\!\pm\!0.08\!\pm\!0.20$GeV$^{-\!1/2}$  \\
3 & $773\pm7$ & $127\pm13$  &  $n\!=\!5.7\!\pm\!0.4\!\pm\!1.5$ \\
\hline
\end{tabular}
\caption{Parameters  for different mass parameterizations.\label{table:fits} }
\vspace*{0.3cm}
\begin{tabular}{|l|c|c|}
\hline 
Cross Section    & STAR (mb) & Ref~\cite{BKN} (mb) \\  
\hline
$\sigma^\rho_{xn,xn}$ & $ 28.3 \!\pm\! 2.0  \!\pm\! 6.3$ &$  27$ \\
$\sigma^\rho_{1n,1n}$ &$ 2.8 \!\pm\! 0.5\!\pm\! 0.7$ & $ 2.6$ \\
\hline
$\sigma^{\rho{\rm (inc. overlap)}}_{xn,xn}$ & $ 39.7  \!\pm\! 2.8  \!\pm\! 9.7$ &$  - $ \\
$\sigma^\rho_{xn,0n}$ & $95 \!\pm\! 60  \!\pm\! 25$ &$  - $ \\
$\sigma^\rho_{0n,0n}$   & $370\!\pm\!170 \pm 80$ & $-$ \\
\hline
$\sigma^\rho_{total}$ & $460\!\pm\!220\!\pm\!110$ & $350$ \\
\hline
\end{tabular}
\caption{Comparison to predictions from~\cite{BKN}. The uncertainties are highly correlated. \label{table}}
\end{table}

For coherent $\rho^0$ production accompanied by mutual nuclear
break-up (xn,xn), we measure a cross section of \xsxnxnpp~in the
two-track event sample, by extrapolating the integral of the
Breit-Wigner fit to full rapidity. By selecting single neutron signals
in both ZDCs, we obtain $\sigma_{1n,1n}^\rho/\sigma_{xn,xn}^\rho \!=\!
0.097\pm0.014$, so
\xssnsn. Single neutron emission is predominantly due to Coulomb
excitation and the subsequent decay of the giant dipole resonance.  The ratio
$\sigma_{1n,1n}^\rho/\sigma_{xn,xn}^\rho$ is consistent with 
$\sigma_{1n,1n}/\sigma_{xn,xn} \!=\! 0.12\pm 0.01$ found for mutual
Coulomb dissociation at RHIC~\cite{mutualbreakup}, supporting that
$\rho^0$ production and nuclear excitation are independent processes.

At $b\!\sim\! 2R_A$ coherent $\rho^0$ photo-production can overlap
with grazing nuclear collisions, producing a low $p_T \; \rho^0$
accompanied by additional tracks. Additional tracks can also be
produced at $b\!>\! 2R_A$ from nuclear excitation by high energy
photons. At present, we can not differentiate between these two
processes.  The coherent (xn,xn) $\rho^0$ sample increases by $40\%$
when events with additional tracks are included. Accounting for this,
we find \xsxnxn. For (1n,1n) events, no additional $\rho^0$ candidates
are found with higher track multiplicities.

The major systematic uncertainties are in the $4\pi$ extrapolation
(15\%), acceptance and reconstruction efficiency (12\%), luminosity
determination (10\%), and event selection (5\%). The overlap region
with grazing nuclear collisions contributes $10\%$; it does not
contribute to $\sigma^\rho_{1n,1n}$, but a $10\%$ uncertainty is due
to the selection of the single neutrons.  These contributions add in
quadrature to $24\%$ systematic uncertainty in the cross sections.

The absolute efficiency of the year 2000 topology trigger is poorly
known and does not allow a direct cross section measurement.  From the
two-track events, we obtain the cross section ratios
$\sigma^\rho_{xn,xn}/\sigma^\rho_{0n,0n} \!=\!  0.09\!\pm\!0.04$ and
$\sigma^\rho_{xn,xn}/\sigma^\rho_{xn,0n} \!=\!0.30 \! \pm \!0.19$.
The uncertainties reflect the small number of (xn,xn) and (xn,0n)
events in the topology trigger data.  Grazing nuclear collisions do
not contribute to $\sigma^\rho_{0n,0n}$ and $\sigma^\rho_{xn,0n}$,
since they yield neutron signals in both ZDCs.  From $\sigma(AuAu
\!\rightarrow\! Au^*_{xn} Au^*_{xn} \rho^0)$, we estimate
\xsnobrk, \xsxnzn, and the total cross section for coherent $\rho^0$ production 
\xstot.
Table~\ref{table} compares our results to the calculations of
Ref.~\cite{BKN}. The calculation for $\sigma^\rho_{xn,xn}$ excludes
grazing nuclear collisions; it is therefore compared to our value
without the overlap correction.  Recent predictions~\cite{Frankfurt}
are about $50\%$ higher than in Ref.~\cite{BKN} without giving specific
numbers for $\sqrt{s_{NN}}=130$~GeV.

In summary, the first measurements of coherent $\rho^0$ production
with and without accompanying nuclear excitation, $AuAu
\!\rightarrow\! Au^\star Au^\star \rho^0$ and $AuAu \!\rightarrow\! Au
Au \rho^0$, confirm the existence of vector meson production in
ultra-peripheral heavy ion collisions.  The $\rho^0$ are produced at
small transverse momentum, showing the coherent coupling to both
nuclei. The cross sections at $\sqrt{s_{NN}}\!=\!130$~GeV are in
agreement with theoretical calculations based on the
Weizs\"acker-Williams approach for large relativistic charges, the
extrapolation from $\rho^0$-nucleon to $\rho^0$-nucleus scattering,
and the assumption that $\rho^0$ production and nuclear
excitation are independent processes.

We thank the RHIC Operations Group and the RHIC Computing Facility
at Brookhaven National Laboratory, and the National Energy Research 
Scientific Computing Center at Lawrence Berkeley National Laboratory
for their support. This work was supported by the Division of Nuclear 
Physics and the Division of High Energy Physics of the Office of Science of 
the U.S. Department of Energy, the United States National Science Foundation,
the Bundesministerium f\"ur Bildung und Forschung of Germany,
the Institut National de la Physique Nucleaire et de la Physique 
des Particules of France, the United Kingdom Engineering and Physical 
Sciences Research Council, Fundacao de Amparo a Pesquisa do Estado de Sao 
Paulo, Brazil, the Russian Ministry of Science and Technology and the
Ministry of Education of China and the National Natural Science Foundation 
of China.

\end{document}